\documentclass[10pt,a4paper]{article}

\usepackage{titlesec}
\titleformat*{\section}{\bf\normalsize\large}
\titleformat*{\subsection}{\bf\normalsize}

\titlespacing{\section}{0pt}{2ex}{1ex}
\titlespacing{\subsection}{0pt}{2ex}{1ex}
\titlespacing{\subsubsection}{0pt}{2ex}{1ex}

\usepackage{fancyhdr}
\usepackage[latin1]{inputenc}
\usepackage[top=2cm,bottom=2cm,left=2cm,right=2cm]{geometry}
\usepackage[usenames,dvipsnames]{color}
\usepackage[pdfborder={0 0 0},colorlinks=true,urlcolor=blue,linkcolor=blue,citecolor=blue,bookmarks=true]{hyperref}
\hypersetup{pdftitle={mlpack~4: a fast, header-only C++ machine learning library}}
\hypersetup{pdfauthor={
Ryan R. Curtin,
Marcus Edel,
Omar Shrit,
Shubham Agrawal,
Suryoday Basak,
James J. Balamuta,
Ryan Birmingham,
Kartik Dutt,
Dirk Eddelbuettel,
Rishabh Garg,
Shikhar Jaiswal,
Aakash Kaushik,
Sangyeon Kim,
Anjishnu Mukherjee,
Nanubala Gnana Sai,
Nippun Sharma,
Yashwant Singh Parihar,
Roshan Swain,
Conrad Sanderson
}}
\usepackage[british]{babel}
\usepackage[none]{hyphenat}
\usepackage{palatino}
\usepackage{inconsolata}

\makeatletter
\renewcommand\@makefntext[1]{%
\noindent
\mbox{\@thefnmark}{#1}}
\makeatother

\newcommand\blfootnote[1]{%
  \begingroup
  \renewcommand\thefootnote{}\footnote{#1}%
  \addtocounter{footnote}{-1}%
  \endgroup
}

\begin{document}

\renewcommand{\baselinestretch}{1.05}\small\normalsize

\thispagestyle{empty}
\pagestyle{empty}

\begin{minipage}{1\textwidth}
\centering

\mbox{\LARGE\bf mlpack~4: a fast, header-only C++ machine learning library}

\vspace{5ex}

Ryan~R.~Curtin,
Marcus~Edel,
Omar~Shrit,
Shubham~Agrawal,
Suryoday~Basak,
James~J.~Balamuta,
Ryan~Birmingham,
Kartik~Dutt,
Dirk~Eddelbuettel,
Rishabh~Garg,
Shikhar~Jaiswal,
Aakash~Kaushik,
Sangyeon~Kim,
Anjishnu~Mukherjee,
Nanubala~Gnana~Sai,
Nippun~Sharma,
Yashwant~Singh~Parihar,
Roshan~Swain,
Conrad~Sanderson
   
\end{minipage}

\vspace{4ex}

\section*{Abstract}

For over 15 years, the mlpack machine learning library has served as a
``swiss army knife'' for C++-based machine learning.
Its efficient implementations of common and cutting-edge machine learning
algorithms have been used in a wide variety of scientific and industrial applications.
This paper overviews mlpack~4, a significant upgrade over its predecessor.
The library has been significantly refactored and redesigned to facilitate
an easier prototyping-to-deployment pipeline, including bindings to other languages
(Python, Julia, R, Go, and the command line)
that allow prototyping to be seamlessly performed in environments other than C++.
mlpack is open-source software, distributed under the permissive 3-clause BSD license;
it can be obtained at
\href{https://mlpack.org}{\texttt{https://mlpack.org}}
\blfootnote{\textbf{Published in:} Journal of Open Source Software, Vol.~8, No.~82, 2023. \textbf{DOI:} \href{https://doi.org/10.21105/joss.05026}{\texttt{10.21105/joss.05026}}}

\vspace{2ex}

\section*{Background}

The use of machine learning has become ubiquitous in almost every scientific
discipline and countless commercial applications~\cite{carleo2019machine,jordan2015machine}.
There is one important commonality to virtually all of these applications:
machine learning is often computationally intensive, due to the
large number of parameters and large amounts of training data.
This was the main motivator for the original development of mlpack in the C++
language, which allows for efficient close-to-the-metal implementations~\cite{curtin2013mlpack,curtin2018mlpack}.

But speed is not everything: development and deployment of applications that use
machine learning can also be significantly hampered if the overall process is
too difficult or unwieldy~\cite{lavin2022technology,paleyes2020challenges}.
Furthermore, deployment environments often have computational or engineering
constraints that make a full-stack Python solution infeasible~\cite{fischer2020ai}.
As such, it is important that lightweight and easy-to-deploy machine learning
solutions are available. This has motivated our refactoring and redesign of
mlpack~4: we pair efficient implementations with easy and lightweight
deployment, making mlpack suitable for a wide range of deployment environments.
A~more complete set of motivations can be found in the mlpack vision document~\cite{mlpack2021vision}.

mlpack is a general-purpose machine learning library, targeting both academic
and commercial use; for instance, data scientists who need efficiency and
ease of deployment, or, e.g., by researchers who need flexibility and
extensibility.  While there are other machine learning libraries intended to be
used from C++, many, such as FAISS~\cite{johnson2019billion} and FLANN
\cite{muja2009fast}, are limited to a few specific algorithms, instead of a full
range of machine learning algorithms, like mlpack provides.  dlib-ml~\cite{dlib09},
on the other hand, does provide a broad toolkit of machine learning algorithms,
but its extensibility is somewhat limited as it does not use policy-based design
\cite{alexandrescu2001modern} to provide arbitrary user-defined behavior, and the
range of machine learning algorithms provided is smaller than mlpack's.

\vspace{1ex}

\section*{Functionality}

The library contains a wide variety of machine learning algorithms,
some of which are new to mlpack~4.  The list of algorithms includes linear regression,
logistic regression, random forests, furthest-neighbor search~\cite{curtin2016fast},
accelerated k-means variants~\cite{curtin2017dual}, kernel density estimation~\cite{lee2008fast},
and fast max-kernel search~\cite{curtin2014dual}.  There is also a module for
deep neural networks, which has implementations of numerous layer types,
activation functions, and reinforcement learning applications.
Details of the available functionality are provided in the online
mlpack documentation (\texttt{https://www.mlpack.org/docs.html}). The efficiency of these
implementations has been shown in various works~\cite{curtin2013mlpack,fang2016m3}
using mlpack's benchmarking system~\cite{edel2014automatic}.

The algorithms are available via automatically-generated bindings to Python,
R, Go, Julia, and the command line. Each of these bindings has a unified interface
across the languages; for example, a model trained in Python can be used from
Julia or C++ (or any other language with mlpack bindings).  The bindings are
available in each language's package manager, as well as system-level package
managers such as \textit{apt} and \textit{dnf}.  Furthermore, ready-to-use Docker containers
with the environment fully configured are available on DockerHub, and an interactive
C++ notebook interface via the \textit{xeus-cling} project (\texttt{https://github.com/QuantStack/xeus-cling})
is available on BinderHub.

Once a user has developed a machine learning workflow in the language of their
choice, deployment is straightforward.  The mlpack library is now header-only,
and directly depends only on three libraries: \textit{Armadillo}~\cite{sanderson2016armadillo},
\textit{ensmallen}~\cite{curtin2021ensmallen}, and \textit{cereal} (\texttt{https://github.com/USCILab/cereal}).
When using C++, the only linking requirement is to an efficient implementation
of BLAS and LAPACK (required via Armadillo).  This significantly eases deployment;
a standalone C++ application with only a BLAS/LAPACK dependency is easily
deployable to many environments, including standard Linux-based Docker containers,
Windows environments, and resource-constrained embedded environments.
To this end, mlpack's build system now also contains a number of tools for
cross-compilation support, including the ability to easily statically link
compiled programs (important for some deployment environments).

\section*{Major Changes}

Below we detail a few of the major changes present in mlpack~4.  For a complete
and exhaustive list (including numerous bug fixes and new techniques), the
\textit{HISTORY.md} file (distributed with mlpack) can be consulted.

\begin{itemize}

\item \textbf{Removed dependencies.}
In accordance with the vision document~\cite{mlpack2021vision}, the majority of the
refactoring and redesign work focused on reducing dependencies and compilation overhead.
This has motivated the replacement of the \textit{Boost C++ libraries} (\texttt{https://www.boost.org}),
upon which mlpack previously depended,
with lightweight alternatives including \textit{cereal} (\texttt{https://github.com/USCILab/cereal})
for serialization.  The entire neural network module was refactored to avoid
the use of \textit{Boost} (amounting to an almost complete rewrite).
This effort was rewarded handsomely: with mlpack~3, a simple program would often
require several gigabytes of memory just for compilation.
After refactoring and removing dependencies, compilation generally requires
just a few hundred megabytes of memory, and is often an order of magnitude faster.

\item \textbf{Interactive notebook environments.}
mlpack can be used in a Jupyter notebook environment~\cite{kluyver2016jupyter}
via the \textit{xeus-cling} project (\texttt{https://github.com/QuantStack/xeus-cling}).
This is demonstrated interactively on the mlpack homepage (\texttt{https://www.mlpack.org}).
Examples of C++ notebooks can be found in the
mlpack examples repository (\texttt{https://github.com/mlpack/examples}),
and these can easily be run on BinderHub.

\item \textbf{New bindings and enhanced availability.}
Support for the Julia~\cite{bezanson2017julia}, Go~\cite{pike2012go}, and R languages
\cite{rcore2022,parihar2022rmlpack} has been added via mlpack's automatic binding
system.  These bindings can be used by installing mlpack from the language's
package manager (\textit{Pkg.jl}, \textit{go get}, \textit{install.packages('mlpack')}).
Furthermore, since mlpack's reduced dependency footprint has significantly
simplified the deployment process, mlpack's Python dependencies are now
available for numerous architectures both on PyPI and in \texttt{conda-forge}.

\item \textbf{Cross-compilation support and build system improvements.}
mlpack's build configuration now supports easy cross-compilation, for instance
via toolchains such as \textit{buildroot} (\texttt{https://buildroot.org}).  By specifying a few
flags, a user may produce a working mlpack setup for a variety of embedded systems.
This required the implementation of a dependency auto-downloader,
which is capable of downloading \textit{OpenBLAS} (\texttt{https://github.com/xianyi/OpenBLAS})
and compiling (if necessary) for the target architecture.  The auto-downloader
can also be enabled and used for any situation, thus easing installation and
deployment.

\end{itemize}

\section*{Acknowledgements}

Development of mlpack is community-led. It is the product of hard work by
over 220 individuals (at the time of writing). We are also indebted to people
that have provided bug reports over the years.  The development has been supported
by Google, via a decade-long participation the Google Summer of Code program,
and also by NumFOCUS, which fiscally sponsors mlpack.

\small
\bibliographystyle{ieee}
\bibliography{references}

\end{document}